 \documentclass[final,3p,times,twocolumn]{elsarticle}
\usepackage{amssymb}
\usepackage{amsmath}
\usepackage{multirow}
\usepackage{dcolumn}
\newcolumntype{.}{D{.}{.}{-1}}

\usepackage{color}

\newcommand{\bma}[1]{\mbox{\boldmath${#1}\/$}}
\newcommand{\Cdot}{\bma{\cdot}}
\newcommand{\Nabla}{\bma{\nabla}}

\journal{Sensors and Actuators A: Physical}

\begin{document}

\begin{frontmatter}


%% \ead{mateos@ice.csic.es}
%% \cortext[cor1]{Corresponding author}

\title{Noise characterization of an atomic magnetometer at sub-millihertz frequencies}


%\corref{cor1}
\author{I. Mateos$^*$}
\ead{mateos@ice.csic.es}
\address{Institut de Ci\`encies de l'Espai (CSIC-IEEC), 08034 Barcelona, Spain}

\author{B. Patton$^1$, E. Zhivun, D. Budker\fnref{fn2}}
%\author[Berkeley]{D. Budker}
\address{Department of Physics, University of California, Berkeley, California 94720-7300}
%\author[Mateos,Lobo]{A. Lobo\fnref{fn1}}
%\fntext[fn1]{Deceased}

\author{D. Wurm}
\address{Technische Universit\"at M\"unchen, 85748 Garching, Germany}

\author{J. Ramos-Castro\fnref{fn3}}
\address{Departament d'Enginyeria Electr\`onica, Universitat Polit\`ecnica de Catalunya, \\08034 Barcelona, Spain}
\cortext[cor1]{Corresponding author}
\fntext[fn1]{Also at Physik-Department, Technische Universit\"at M\"unchen, 85748 Garching, Germany}
\fntext[fn2]{Also at Lawrence Berkeley National Laboratory, Berkeley, California 94720 and Helmholtz Institute, Johannes Gutenberg University, 55099 Mainz, Germany}
\fntext[fn3]{Also at Institut d'Estudis Espacials de Catalunya, 08034, Barcelona, Spain}

%\address[Ramos]{Departament d'Enginyeria Electr\`onica, Universitat Polit\`ecnica de Catalunya, 08034 Barcelona, Spain}

% \author{I. Mateos}
% \email{mateos@ice.csic.es}
% \affiliation{Institut de Ci\`encies de l'Espai (CSIC-IEEC), 08034 Barcelona, Spain}
% \author{B. Patton}\altaffiliation[Also at ]{Physik-Department, Technische Universit\"at M\"unchen, 85748 Garching, Germany}
% \author{E. Zhivun}
% \author{D. Budker}\altaffiliation[Also at ]{Lawrence Berkeley National Laboratory, Berkeley , California 94720 and Helmholtz Institute, Johannes Gutenberg University, 55099 Mainz, Germany}
% \affiliation{Department of Physics, University of California, Berkeley, California 94720-7300}
% 
% \author{D. Wurm}
% \affiliation{Technische Universit\"at M\"unchen, 85748 Garching, Germany}
% \author{J. Ramos-Castro}\altaffiliation[Also at ]{Institut d'Estudis Espacials de Catalunya, 08034, Barcelona, Spain}
% \affiliation{Departament d'Enginyeria Electr\`onica, Universitat Polit\`ecnica de Catalunya, \\08034 Barcelona, Spain}

\begin{abstract}
% Noise measurements have been carried out  in the LISA bandwidth (0.1\,mHz to 100\,mHz) to characterize an all-optical atomic magnetometer based on nonlinear magneto-optical rotation. This was done in order to assess if the technology can be used for space missions with demanding low-frequency requirements like the LISA concept. Magnetometry for low-frequency applications is usually limited by intrinsic $1/f$ noise and thermal drifts, which become the dominant contributions at sub-millihertz frequencies. Magnetic field measurements with atomic magnetometers are not immune to low-frequency fluctuations and significant excess noise may arise due to external elements, such as drifts in the applied magnetic field and intrinsic noise or temperature fluctuations in the electronics. We have found the technology suitable for LISA in terms of sensitivity, although further work must be done to characterize the low-frequency noise in a miniaturized setup suitable for space missions.

Noise measurements have been carried out  in the LISA bandwidth (0.1\,mHz to 100\,mHz) to characterize an all-optical atomic magnetometer based on nonlinear magneto-optical rotation. This was done in order to assess if the technology can be used for space missions with demanding low-frequency requirements like the LISA concept. Magnetometry for low-frequency applications is usually limited by $1/f$ noise and thermal drifts, which become the dominant contributions at sub-millihertz frequencies. Magnetic field measurements with atomic magnetometers are not immune to low-frequency fluctuations and significant excess noise may arise due to external elements, such as  temperature fluctuations or intrinsic noise in the electronics. In addition, low-frequency drifts in the applied magnetic field have been identified in order to distinguish their noise contribution from that of the sensor. We have found the technology suitable for LISA in terms of sensitivity, although further work must be done to characterize the low-frequency noise in a miniaturized setup suitable for space missions.

%Although drifts in the applied magnetic field are not considered as magnetometer noise, its characterization is crucial to discern their influence at low frequency.
\end{abstract}


\end{frontmatter}



\section{Introduction} \label{introduction}

\begin{figure}[ht!]
\centering
\centering
\includegraphics[width=1.00\columnwidth]{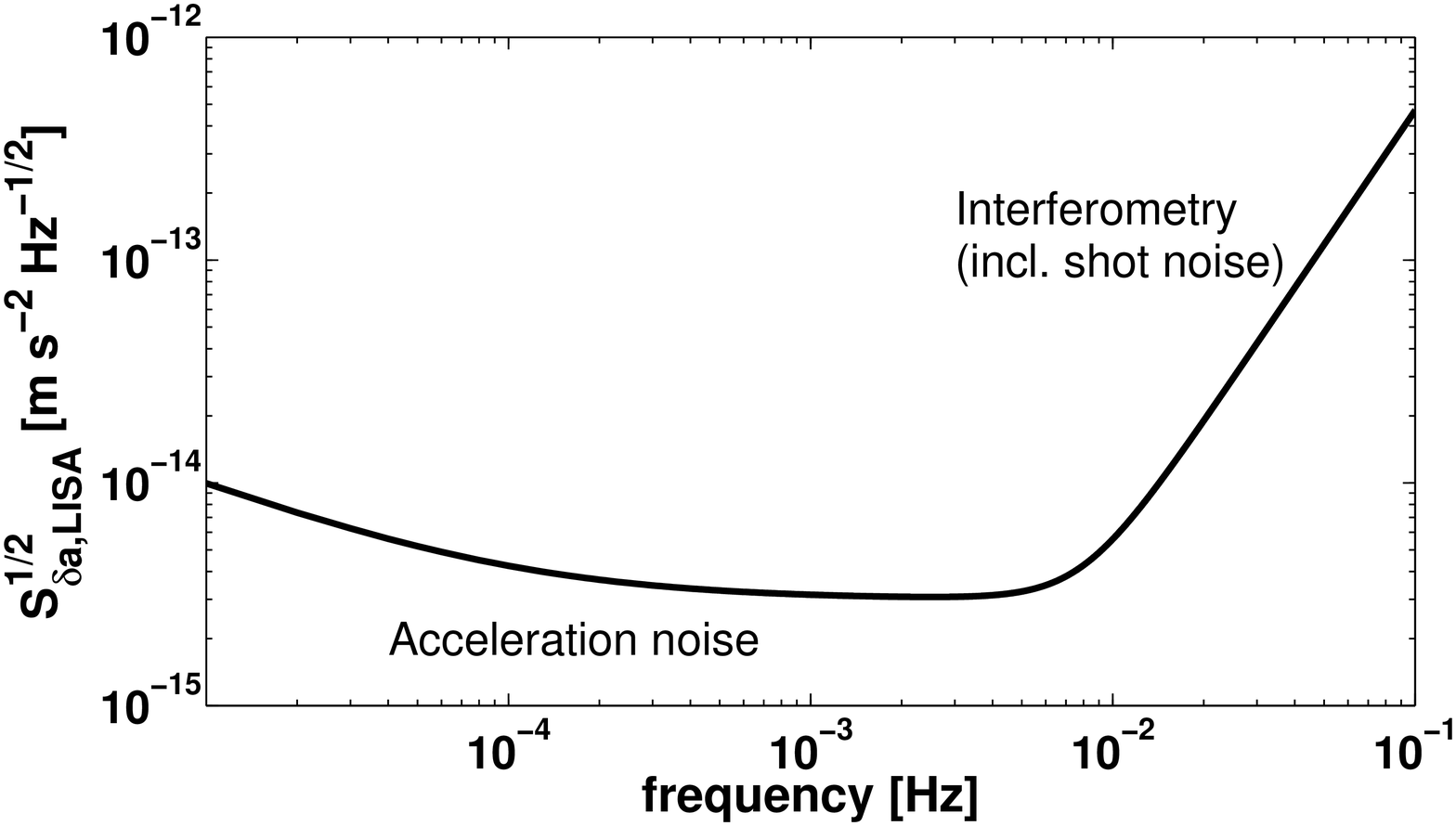}
\caption{LISA noise requirement plotted as amplitude spectral density (ASD) of the differential acceleration of the test masses. \label{fig:req}}
\end{figure}

The evolved Laser Interferometer Space Antenna (eLISA) is a mission concept proposed as a large (L-class) space mission of the European Space Agency (ESA) designed to detect low-frequency gravitational radiation. It will be formed by three drag-free spacecraft in triangular configuration with one-million-kilometer arms, with each spacecraft containing one or two ``free-falling'' macroscopic bodies called test masses (TMs) \cite{whitepaper}. Gravitational wave (GW) detection requires interferometry measurement at the picometer level between two test masses along an arm due to the tidal deformation of spacetime caused by GWs. For this reason, the environment where such bodies will be located must be very quiet (in relation to disturbances exerting forces in the bodies), otherwise the motion provoked by the different noise sources perturbing the free floating bodies would conceal the GW signal. The LISA top-level requirement  in terms of acceleration noise $S^{1/2}_{\delta a, {\rm LISA}}$ (${\rm m \ s^{-2} \ Hz^{-1/2}}$) in the frequency band between $0.1\,{\rm mHz}\,\leq \omega/2\pi \leq\,100\,{\rm mHz}$ is plotted in Fig.\,\ref{fig:req}. At frequencies below 1 mHz, the noise is dominated by the residual acceleration noise of $\sqrt{2}\cdot3\,{\rm fm\,s^{-2} Hz^{-1/2}}$ caused by magnetic or temperature forces among others \cite{bib:LISAReq}. At higher frequencies, the sensitivity is limited by the arm-length-measurement noise in the interferometer \cite{bib:LISAReqII}.

Owing to the magnetic properties of the TMs, more precisely their magnetization ${\bf M}$ and magnetic susceptibility $\chi$, one of the main contributions to the total acceleration noise budget is the magnetic field ${\bf B}$ in the spacecraft. The force on the TM volume $V$ induced by a magnetic disturbance is given by

\begin{equation}\label{equ:MagForce}
 {\bf F} = \left\langle\left[\left({\bf M} + 
           \frac{\chi}{\mu_0}\,{\bf B}\right)\Cdot\Nabla\right]{\bf B}
           \right\rangle V. 
\end{equation}

%To achieve maximum efficiency of the instrument
Since the magnetic background in the spacecraft can induce a residual acceleration noise on the TMs and therefore deteriorate the efficiency of the instrument, its magnetic contribution needs to be quantified and suppressed from the main data stream. For this reason, magnetic sensors are needed to map the environmental magnetic field and its gradient. These magnetometers can not be placed at the sites of the TMs, which is the region of interest, in consequence an interpolation method needs to be implemented\,\cite{bib:MagInter,bib:MagInterII}.

The low-frequency noise limits the performance of the on-board instrumentation. Hence, magnetometers and optimized electronics need to be assessed in the LISA bandwidth, since at sub-mHz frequencies, sensor noise is usually dominated by the $1/f$ contribution and thermal drifts. Previous studies on this subject were carried out with magnetoresistance-based magnetometers, suggesting the technology as a potential option for eLISA\,\cite{bib:AMRLockin}. We are currently investigating sensors based on alkali-vapor cells  \cite{BudkerBook} as an alternative to fluxgate magnetometers, which are commonly employed in space applications.  Alkali-vapor magnetometers are absolute sensors, since the spin-precession frequency is related to the magnitude of the field by fundamental constants. In this regard, they are unlike fluxgates and magnetoresistances, which measure small changes in the field but not its actual value without precise calibration. Therefore, it is plausible that atomic magnetometers might be better for low-frequency applications. Theoretically, their sensing elements are not subject to intrinsic $1/f$ noise, although in practical situations, there are noise sources that might be relevant at very low frequencies. For instance, drifts in the power, wavelength, or polarization of the pump and probe lasers will determine the ``light shift''; a light-induced shift in the alkali Zeeman energy sublevels which can mimic the effects of an applied magnetic field\,\cite{bib:Mathur,bib:Cohen}. The overall light shift will also be influenced by changing cell temperature or alkali-vapor density, which will affect attenuation of the beams within the vapor. Other potential sources are the drifts in the phase or duty cycle of the pump laser waveform, intrinsic noise and temperature dependence of the electronics and changes in the stray magnetic field from the components surrounding the vapor cell. By definition, the latter is not categorized as magnetometer noise.  Nevertheless, it is crucial to disentangle the influence of magnetic-field drifts so that we may study the intrinsic sensor noise. It is important to remark that eLISA is a demanding mission in terms of low-noise/low-frequency concepts. For these reasons, the study below the corner frequency of the $1/f$ noise is critical and it differs from the wideband applications, where usually only the noise floors are of concern. Some of the estimates we performed are based on the $1/f$ behavior, which is characterized by the  white-noise floor and the corner frequency. Other parameters also critical in the low-frequency band, such as ambient temperature fluctuations, are derived from laboratory measurements.

% \begin{itemize}
% 
%  \item drifts in the power, wavelength, or polarization of the pump and probe lasers. These parameters determine the ``light shift'', a light-induced shift in the alkali Zeeman energy sublevels which can mimic the effects of an applied magnetic field,\cite{bib:Mathur,bib:Cohen}
%  \item shifts due to changing cell temperature or alkali-vapor density. Alkali density fluctuations will affect attenuation of the beams within the vapor, affecting the overall light shift,
%  \item drifts in the phase or duty cycle of the pump laser waveform,
%  \item intrinsic noise and temperature dependence of the electronics,
%  \item changes in the magnetic field surrounding the vapor cell. 
%  
%  
% \end{itemize}

\subsection{Atomic magnetometry: Preferred techniques}

Atom-based magnetometers are the most sensitive devices to measure magnetic fields, furthermore they do not need the bulky and expensive cryogenic refrigeration required in superconducting quantum interference devices (SQUIDs).  The {\em spin-projection-noise-limited} sensitivity $\delta B_{\rm SNL}$ of an atomic magnetometer during a measurement period $T$ is given by
\begin{equation}
 \delta B_{\rm SNL} \simeq \frac{1}{\gamma \sqrt{N \,\tau\, T}},
\end{equation}

\noindent where $\gamma$ is the gyromagnetic ratio of the atoms and $\delta B_{\rm SNL}$ scales with the size and temperature of the cell by means of the coherence time $\tau$ and the total number of atoms $N$.  Since the fundamental sensitivity limit of atomic magnetometers is much better than the required sensitivity for eLISA, we can trade off sensitivity for size/temperature of the cell, motivating the technology for space missions. Previous results assuming the spin-projection noise as the limiting noise source, suggest that magnetometer containing optimized cells with a size between $1\,{\rm cm}$ and $10\,{\rm \mu m}$ can reach noise levels around $1 \,{\rm fT \,Hz}^{-1/2}$ and $10 \,{\rm pT \,Hz}^{-1/2}$, respectively. The corresponding cell temperature may vary from room temperature for the largest cell to $110^{\rm o}{\rm C}$ for the $10\,{\rm \mu m}$ cell\,\cite{bib:Shah}.

Various techniques to measure the Larmor precession frequency $\Omega_{\rm L}$ of atomic spins have achieved excellent sensitivities at room temperature, although most of these techniques have been studied at higher frequencies ($\geq0.1\,{\rm Hz}$). Our interest consists in the detection of small magnetic fluctuations in the low-frequency region and at ambient temperature. Hence, based on the experience with the Magnetic Diagnostics for LISA Pathfinder (eLISA's precursor mission) constituted by a set of four tri-axial fluxgate magnetometers and two coils\,\cite{LPFMag}, the main sensor selection criteria for eLISA are viability for miniaturization, low noise in the millihertz region and small back-action effect on the spacecraft environment. This is due to, firstly, the fact that space applications have strict requirements in size and weight, and smaller sensors allow more of them to be incorporated in the spacecraft. Furthermore, spatial resolution is increased with a more compact sensor head. Secondly, regarding the sensor noise and frequency range, in the more demanding scenario both shall be one order of magnitude lower than the LISA Pathfinder requirement ($10 \,{\rm nT \,Hz}^{-1/2}$ at $1\,{\rm mHz}$). This implies a noise level in the measurement system of

 \begin{equation}
 S^{1/2}_{B,\rm system} \leq 1 \,{\rm nT \,Hz}^{-1/2},\,\,0.1\,{\rm mHz}\leq\frac{\omega}{2\pi}\leq 100\,{\rm mHz}.
 \label{eq:req}
 \end{equation}

 \noindent Finally, the selected magnetometer should have sufficiently low magnetic and thermal back-action effects on the spacecraft environment to avoid disturbances on the TMs \cite{MateosBack}. A sensor fulfilling these requirements would also be well-suited for use in magnetically sensitive fundamental physics experiments requiring long integration time, such as the search for a permanent electric dipole moment of the neutron \cite{bib:nedm} and high-precision measurements of the weak equivalence principle using space atom interferometry in STE-QUEST \cite{bib:ste-quest}.

 %Magnetic and thermal environmental conditions are not fully defined for eLISA, but at least they should meet the LISA Pathfinder estimations, i.e., $100\,{\rm nT/Hz^{-1/2}}$ and $250\sqrt{3}\,{\rm nT m^{-1}/Hz^{-1/2}}$ for the magnetic field and its gradient and $100\,{\rm \mu K/Hz^{-1/2}}$ for the temperature fluctuations, both at 1 mHz\,\cite{MateosBack,bib:LoboLISA7}.
 
% \begin{itemize}
%  \item viability for miniaturization: Space applications have strict requirements in size/weight, and smaller sensors allow more of them to be incorporated in the spacecraft. Spatial resolution is also increased with a more compact sensor head;
%  \item low noise in the millihertz region: In the more demanding scenario it shall be one order of magnitude below in amplitude and frequency than the LISA Pathfinder requirement ($10 \,{\rm nT \,Hz}^{-1/2}$ at $1\,{\rm mHz}$). It implies a sensitivity in the measurement system of 
%  
%  \begin{equation}
%  S^{1/2}_{B,\rm system} \leq 1 \,{\rm nT \,Hz}^{-1/2},\,\,\omega/2\pi = 0.1\,{\rm mHz};
%  \label{eq:req}
%  \end{equation}
%  
%  
%  \item magnetometers with sufficiently low magnetic and thermal back-action effects on the spacecraft environment to avoid disturbances on the TMs\,\cite{MateosBack}.
% \end{itemize}

\begin{figure*}[bt]
 \includegraphics[width=0.95\textwidth]{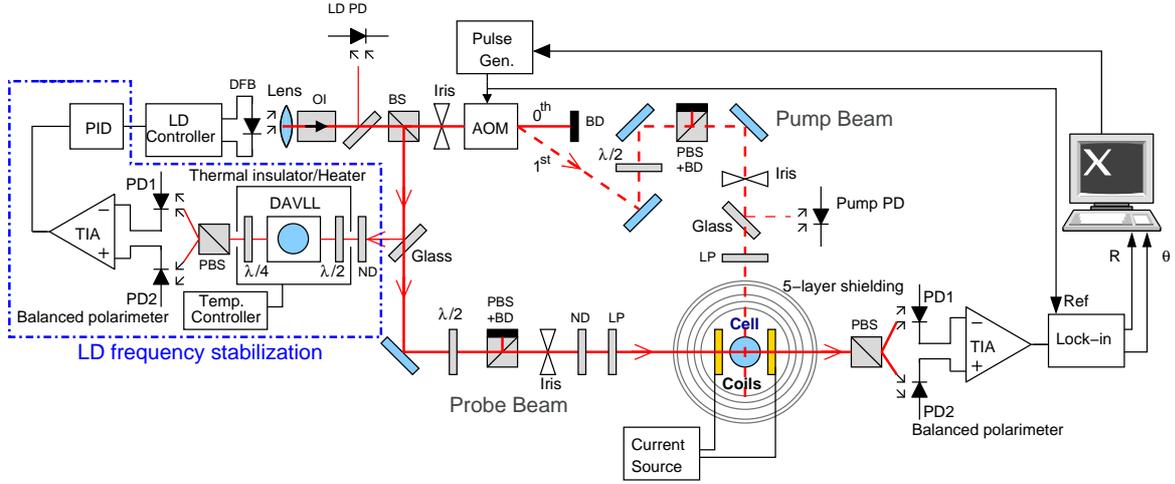}
 \caption{Experimental setup. DFB: distributed feedback laser, OI: optical isolator, (P)BS: (polarizing) beam splitter, AOM: acousto-optic modulator, BD: beam dump, $\lambda/2$: half-wave plate , $\lambda/4$: quarter-wave plate, PD: photodiode, LP: linear polarizer, ND: neutral density filter, DAVLL: dichroic-atomic-vapor laser lock, TIA: transimpedance amplifier. \label{fig:Setup}}
\end{figure*} 

Chip-scale magnetometers based on magnetic-resonance phenomena can be driven either with radio frequency (RF) fields or with modulated light. A coil-driven magnetometer\,\cite{atomMagMx} with micro-fabricated vapor cell has demonstrated noise levels of $5\,{\rm pT\,Hz}^{-1/2}$ for a bandwidth from 1 to 100\,Hz, however the magnetic field created by the RF coils for this method could constitute a potential source of disturbances to the eLISA performance \cite{MateosBack}. Similar noise level has been measured in a frequency-modulated Bell-Bloom magnetometer (FM BB), using also millimeter-scale cells \cite{MxandFMBB,NMORNoHeaters}. For the two previous arrangements the cell needs to be heated to create sufficient atomic density for the measurement, which could also be detrimental for the quiet thermal environment required in the TM region ($ \leq 100\,{\rm \mu K\,Hz^{-1/2}}$ at $1\,{\rm mHz}$ for LISA Pathfinder) \cite{Gibert}. In this work we study the low-frequency noise in a magnetometer prototype based on  nonlinear magneto-optical rotation (NMOR) \cite{FMNMOR,NMOR}, which retains the all-optical excitation with the advantage that the vapor cell is at room temperature and heaters are not utilized \cite{NMORNoHeaters}. In Sec.\,\ref{sec:set-up} we discuss the test setup using nonlinear magneto-optical rotation with amplitude-modulated light (AM-NMOR). In Sec.\,\ref{sec:electronic} we analyze the noise contributions of two important circuits for the magnetic field measurements in the experiment. The experimental results  are shown in Sec.\,\ref{sec:results} and, finally, the main conclusions of the work are summarized in Sec.\,\ref{sec:conclusions}.

\section{Magnetometer setup using AM-NMOR}\label{sec:set-up}

The sensor core is a 1-cm diameter and 3-cm long cylindrical antirelaxation-coated cell containing $^{133}{\rm Cs}$ atoms. The light of a distributed-feedback (DFB) laser \cite{DFB853} is split in a two-beam arrangement with linearly polarized probe and pump light, and it is frequency stabilized near the ${\rm D}_2$ line by means of a dichroic-atomic-vapor laser lock (DAVLL) \cite{DAVLL,DAVLL2}.  The linearly polarized pump beam is square-wave modulated with a $10\%$ duty cycle at a frequency of $\sim2\Omega_{\rm L}$ in order to generate atomic alignment \cite{alignment}. This amplitude modulation is provided by an acousto-optic modulator (AOM) with a RF of $80\,{\rm MHz}$ such that it drives the coherent precession of the atoms about the external magnetic field to be measured \cite{2wl}. The continuous probe and modulated pump beam pass through the vapor cell with approximately the same time-averaged light power of $\sim 1\,{\rm \mu W}$.  Finally, the amplitude of the probe optical rotation is measured with a balanced polarimeter at the modulation frequency $\Omega_m$, where the current difference between the two silicon photodiodes (OSD 15-0) is changed to voltage by a transimpedance amplifier (TIA) and is then demodulated with a phase-sensitive detector. The experimental schematic is shown in Fig.\,\ref{fig:Setup}. 

\begin{figure}
\centering
\includegraphics[width=1.0\columnwidth]{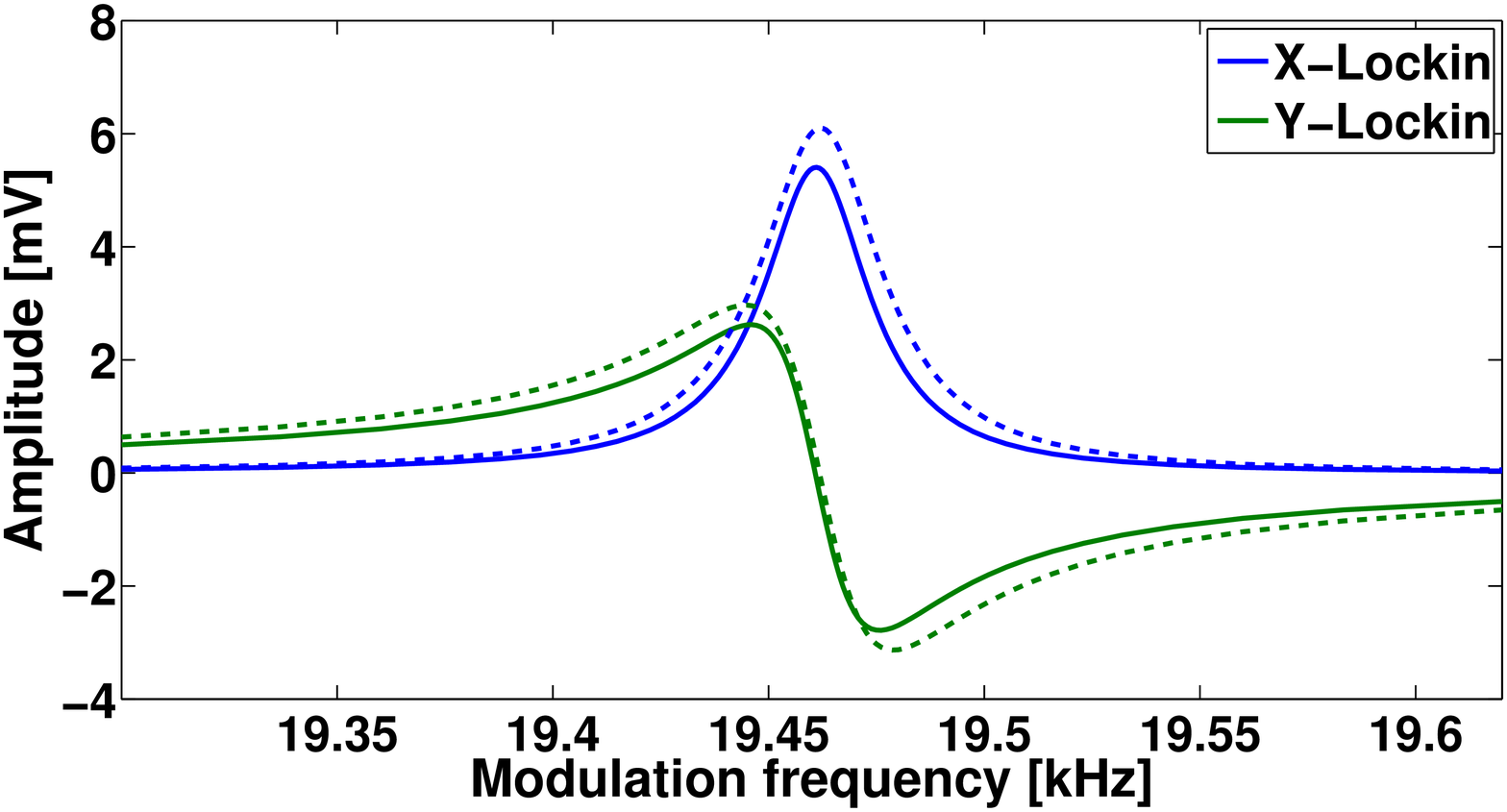}
\includegraphics[width=1.0\columnwidth]{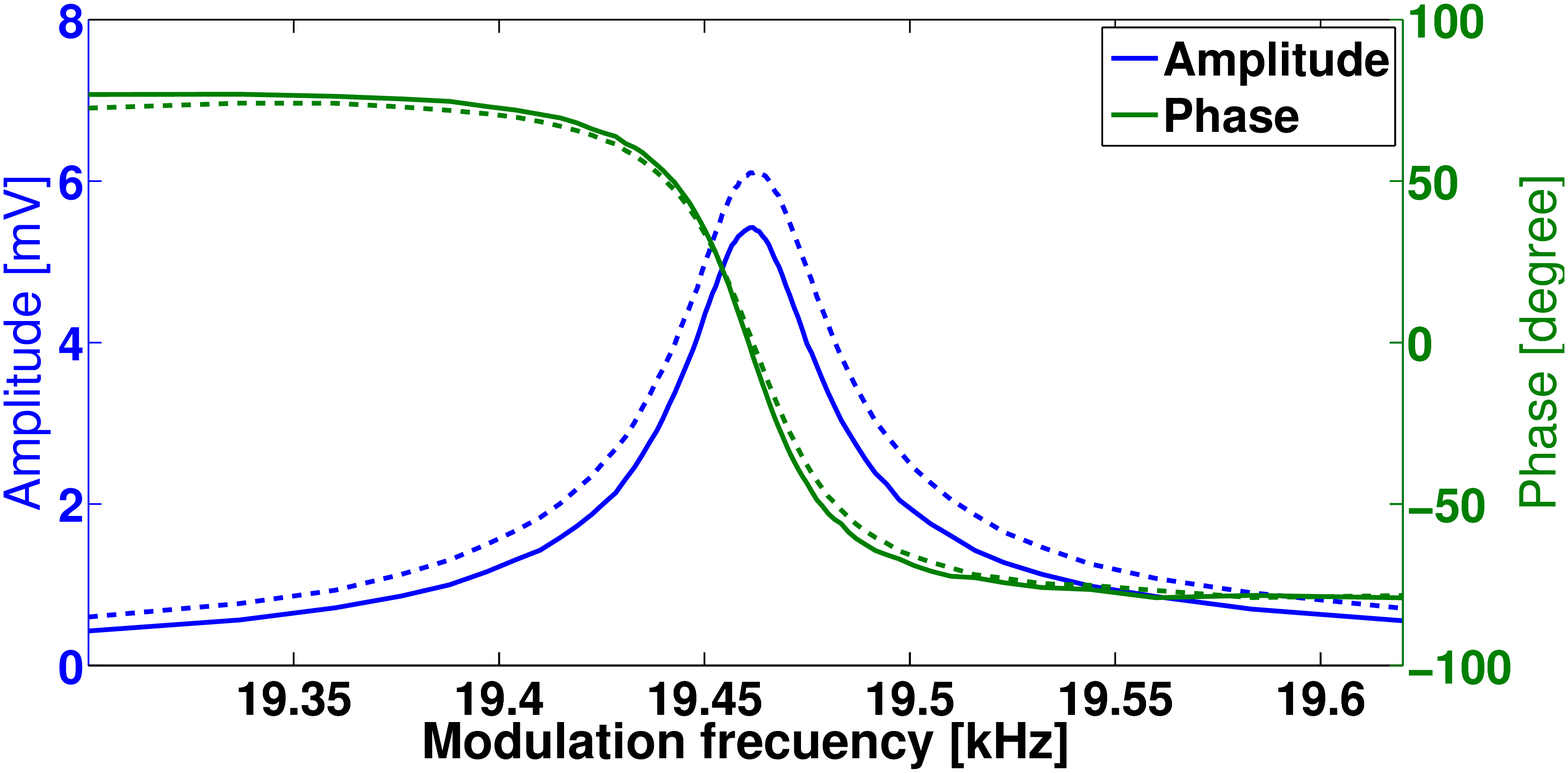}
\caption{Changes in the resonance curve during the long-term measurements. The plots show the outputs of the lock-in amplifier as a function of the modulation frequency, before (dashed line) and after (solid line) a 12-hours run. Traces show Lorentzian fits to the data. Top: Absorptive ($X$) and dispersive ($Y$) components of the magnetic-resonance signal. Bottom: Equivalent amplitude and phase.  \label{fig:FO}}
\end{figure}

There are different ways to measure the magnetic-resonance frequency with the same setup:  We can map out the whole resonance curve by stepping the modulation frequency, or we can make a single-point measurement of the dispersive trace near the center of the resonance. For the former open-loop method, the dynamic range and the measurement rate are limited by the narrow resonance and the slow scan of the resonance curve, respectively. The latter method requires a continuous closed-loop mode to keep the in-phase component locked, i.e., a digital controller that follows the null output of the quadrature component or the equivalent phase signal by tuning the frequency of the pulse generator which drives the modulation. This method of magnetometer operation is also useful to track slow drifts in the measured magnetic field. As can be seen in Fig.\,\ref{fig:FO} during long-term measurements ($\sim12$ hours), the magnetic-resonance signal used to measure the Larmor frequency shows variations in the amplitude of the in-phase (absorptive) component, as well as the quadrature (dispersive) component  obtained from the lock-in amplifier output. Barring fluctuations in the phase of the resonance, the zero-crossing of the dispersive term (at $\Omega_m = 2\Omega_{\rm L}$) does not shift when either the amplitude or the width of the resonance changes. Such changes can arise from variations in the cell temperature or laser power.

For noise measurements, the Cs cell is placed inside a five-layer $\mu$-metal shielding equipped with a solenoidal coil to apply a bias field along the probe-beam path.  An external field is applied to operate the magnetometer at a frequency higher than the linewidth of the resonance, where synchronous optical pumping is employed. Magnetic field fluctuations inside the shield can be caused by the thermal magnetization noise, Johnson noise currents within the magnetic shield itself and unshielded ambient magnetic field fluctuations, but they are not expected to be the dominant source in our measurement ($< 100\,{\rm fT\,Hz^{-1/2}}$)\,\cite{BudkerBook,shieldI,shieldII,shieldIII}. One of the critical parts of the setup in the low-frequency band is the stability of the magnetic field created by the coils, which is mostly a requirement on the coil current source. Hence, special care needs to be taken in the design of the current source to avoid it becoming the dominant noise contribution in the millihertz regime and obscuring the intrinsic noise of the sensor. 

In order to address the equivalent magnetic field noise of the magnetometer we record the setup parameters, namely power, wavelength, current and temperature of the laser diode, power of the pump light, current through the coil, room temperature and magnetic field, polarimeter outputs (single-sided and differential) and lock-in amplifier outputs. These measurements help us disentangle the different contributions to study the magnetometer noise, e.g., how the temperature drifts can affect the electronics.

The setup contains table-top optics in order to facilitate the optimization of the different parts and parameters of the experiment. However, future work will include efforts to miniaturize the magnetometer. The miniaturized design can be based on fully integrated chip-scale magnetometers\,\cite{atomMagMx} or microfabricated remote sensor heads coupled to the laser and photodiodes through optical fibers\,\cite{BudkerBook}. The evident benefit of the first design is the possibility to include all the components in a single chip. In contrast, the advantage of the latter approach is to keep the cell clear of possible magnetic disturbances caused by the proximity of the laser and electrical connections. Besides, the second design is specially useful when an array of sensors is required since some parts can be shared, as for example the stabilized laser, AOM and optical elements.

\section{Electronic noise contributions}\label{sec:electronic}

As mentioned in Sec.\,\ref{introduction}, there are many potential sources of magnetic-resonance frequency drift and intrinsic noise in the magnetometer setup.  In this analysis we focus on two circuits that need to be carefully designed in order to minimize their contribution to the resultant total noise, these circuits have been analyzed at two different frequencies. At millihertz scales, an important electronic noise contribution might be attributed to the leading-field current source.  At higher frequencies $2\Omega_{\rm L}$, the noise of the probe-beam polarimeter can dominate the noise floor; which is relevant to making measurements with high signal-to-noise ratio. The DAVLL also uses a balanced polarimeter to provide an error signal which locks the laser to the atomic resonance.  This signal is DC, so slow drifts in the electronic output of the DAVLL polarimeter can cause slow wavelength fluctuations in the laser, thus creating a time-varying light shift within the vapor.  We are presently investigating this potential source of magnetometer noise but have reason to believe that it is small compared to the noise contributions studied here because the linearly polarized pump beam contributes a negligible systematic shift to the magnetic-resonance frequency \cite{bib:Jensen}. Other than the two aforementioned circuits the rest of the electronic boxes shown in Fig.\,\ref{fig:Setup} have been selected from commercial instrumentation.

\subsection{Equivalent Magnetic field noise due to the leading-field current source}\label{sec:noiseFL}
  
\begin{figure}[ht!]
 \begin{center}
  \includegraphics[width=1.0\columnwidth]{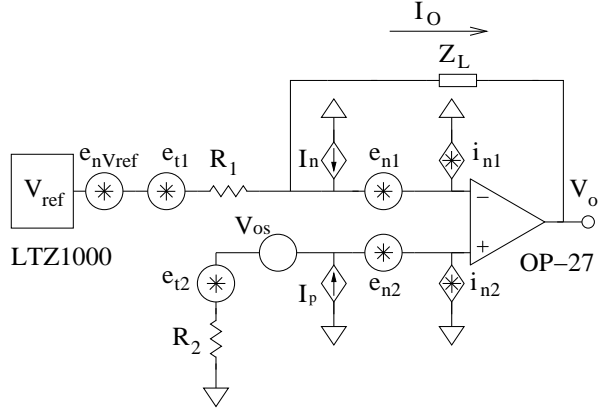}
 \end{center}
\caption{{\em Floating-load} current source with the main sources considered for the noise estimation. Noise parameters for the op-amp and voltage reference are shown in Table\,\ref{tab:componentes}. Manufactures specify $e^{2}_{\rm n} = e^{2}_{\rm n1} + e^{2}_{\rm n2}$ and  $i_{\rm n} = i_{\rm n1} = i_{\rm n2} $.}
\label{fig:load_loop_noise}
\end{figure}

Among the different coil current source topologies that have been analyzed for the experiment, the {\em floating-load} current source shown in Fig.\,\ref{fig:load_loop_noise} has been selected on the grounds of its slightly lower noise and better thermal performance. For the analysis, the noise sources that have been identified in the electronics are the input noise of the operational amplifier, the Johnson noise and the temperature coefficient (TC) in the resistors, as well as thermal dependences of the operational amplifier parameters. The latter are composed of the thermal drift in the bias current TC($I_B$), offset current TC($I_{OS}$) and offset voltage TC($V_{OS}$) of the operational amplifier, and they can be neglected since their overall effect is much smaller ($0.03\,{\rm pA\,K^{-1}}$) than the thermal effect of the resistance ($4.3\,{\rm nA\,K^{-1}}$). Hence, the overall power spectral density of the current source $S_{I_{\rm o}}$ is approximated as

\begin{align}
 \nonumber
 S_{I_{\rm o}}(\omega) & = S_{I_{\rm o, noise}}(\omega)  + \left(\frac{\partial Io}{\partial T}\right)^{2}S_{T}(\omega)\\
 & \simeq i_{\rm n}^2 + \frac{1}{R_1^2}\left(e_{\rm n}^2 + 4k_{\rm B}TR_{1} +   e_{\rm n,Vref}^2 + V_{\rm ref}^2 \alpha^2 S_{T} \right), \label{eq:Si}
\end{align}

\noindent where $i_{\rm n}$ and  $e_{\rm n}$ is the op-amp current and voltage noise spectral density, $4k_{\rm B}R_{1}T$ is the Johnson noise component, $k_{\rm B}$ is the Boltzmann constant, $R_1 = 1\,{\rm k\Omega}$ is the current source resistance, $T$ is the temperature, $e_{\rm n,Vref}$ is the voltage noise of the voltage reference, $V_{\rm ref}$ is the output of the voltage reference, $\alpha = 0.6\,{\rm ppm\,K^{-1}}$ is the TC of the resistors  and  $S_{T}$ is the room temperature fluctuations in power spectral density. All the terms in Eq.\,\eqref{eq:Si} are frequency dependent except the Johnson noise.  Low-frequency noise in the voltage reference and operational amplifier was modeled by the corner frequency at which $1/f$ noise matches the white noise. We used the corner frequency and spectral densities given by the manufacturer, or by experimental data fit. Table\,\ref{tab:componentes} gives output noise parameters for the op-amp and voltage reference used in the current source.

\begin{center}
\begin{table}[ht!]
\caption{Output noise parameters for the components used in the current source. Voltage reference (VR) is based on the LTZ1000 Zener reference.}
\begin{center}
%\begin{ruledtabular}
\begin{tabular}{c c . c c}
\hline\hline
\multirow{2}{*}{IC} & \multicolumn{1} {c}{$e_n$} & \multicolumn{1} {c}{$f_{c,e_n}$}   & \multicolumn{1} {c}{$i_n$} & \multicolumn{1} {c}{$f_{c,i_n}$}\\
            & \multicolumn{1} {c}{$[{\rm  nV/\sqrt{\rm Hz}}]$} & \multicolumn{1} {c}{$[{\rm Hz}]$} & \multicolumn{1} {c}{$[{\rm pA/\sqrt{\rm Hz}}$]} & \multicolumn{1} {c}{$[{\rm Hz}]$}\\\hline
OP27  & 3 & 2.7 & 0.4 & 140\\ 
VR    & 46  & 30 & -& -\\\hline\hline
\end{tabular}
%\end{ruledtabular}
\end{center}

\label{tab:componentes}
\end{table}
\end{center}

\begin{figure}[ht!]
\centering
  \includegraphics[width=1.00\columnwidth]{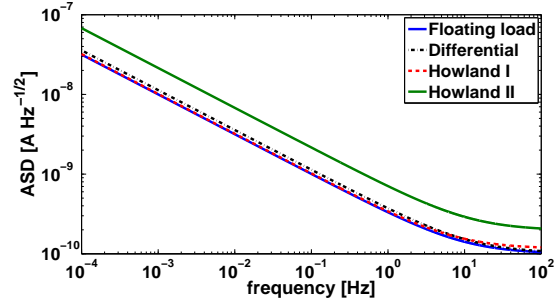}
\caption{Current spectral density for {\em floating-load}, {\em differential}, {\em classical Howland} and {\em improved Howland}  current sources. The voltage reference dominates the noise in the measurement bandwidth.}
\label{fig:CurrentSourceAnalysis}
\end{figure}

\begin{figure}[ht!]
\centering
  \includegraphics[width=1.00\columnwidth]{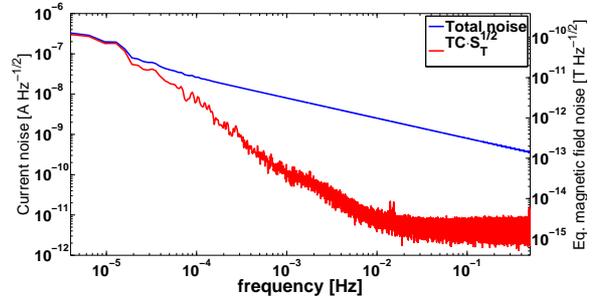}
\caption{Current and equivalent magnetic field spectral densities for the \textit{floating-load} current source (blue) and thermal contribution due to the current source's TC (red). Equivalent magnetic field noise is obtained for a current-to-field ratio of $\sim 392\,{\rm nT/mA}$.}
\label{fig:result_floating}
\end{figure}

A current of $7.1\,{\rm mA}$ is sent through the coil, producing a leading magnetic field of $2.8\,{\rm \mu T}$. For this value of load, Fig.\,\ref{fig:CurrentSourceAnalysis} shows the estimated noise density for the selected {\em floating-load} source in comparison with three other classical topologies \cite{franco}. Fig.\,\ref{fig:result_floating} shows the calculated noise densities obtained for the selected current source, including the contribution due to the thermal fluctuations measured in the laboratory. In the figure, the excess noise due to the thermal dependences of the circuit is significant only below $0.1\,{\rm mHz}$, and thus is outside the eLISA bandwidth. The equivalent magnetic field noise is found by multiplying the current noise expressed in  Eq.\,\eqref{eq:Si} by the current-to-field conversion of the coil.

The theoretical noise  is $ 25 \,{\rm nA\,{Hz}^{-1/2}}$ at $0.1\,{\rm mHz}$, which corresponds to an equivalent magnetic field noise of $ 10\,{\rm pT\,{Hz}^{-1/2}}$. This is well below the more demanding scenario for the magnetometer noise level in Eq.\,\eqref{eq:req}. The result implies that the designed current source achieves the performance required for the noise measurements of $0.1\,{\rm nT\,{Hz}^{-1/2}}$ at $0.1\,{\rm mHz}$, i.e., to be on the safe side, stability of the applied magnetic field must be at least one order of magnitude less noisy than the limit imposed by the magnetometer requirement or the expected noise of the magnetometer under study. The main source of technical noise in the whole bandwidth is the voltage reference, hence, making use of low-noise voltage references \cite{vref} or batteries \cite{battery} will help to improve the noise performance significantly. By eliminating this technical noise, the equivalent magnetic field noise would be reduced to around $6.3\,{\rm nA\,Hz^{-1/2}}\,(2.5\,{\rm pT\,Hz^{-1/2}})$ at $0.1\,{\rm mHz}$ without any thermal insulator or active temperature controller.

\subsection{Polarimeter noise analysis}\label{sec:NoisePol}

The polarimeter circuit is a two-stage amplifier formed by a conventional TIA topology and a non-inverting amplifier in the second stage. The differential signal resulting from the rotation of the light polarization, that is,  the difference of the current from two photodiodes, is converted to voltage with a sensitivity of $11\,{\rm mV/nA}$. The differential photocurrent induced by the probe rotation is modulated at $2\Omega_{\rm L}$, so it is the noise floor of the polarimeter at this frequency which determines the magnetometer's sensitivity.  This frequency can range from the $\sim$Hz scale to hundreds of kHz at Earth's field. For the eLISA mission the magnetic-resonance frequency will possibly be around hundreds of Hz, although the information available so far is not definitive and a wider range needs to be considered.  In any case, the low-frequency analysis is important to study the corner frequency of the $1/f$ noise, since it could be within the bandwidth of modulation.

\begin{figure}[ht!]
\centering
 \includegraphics[width=1.03\columnwidth]{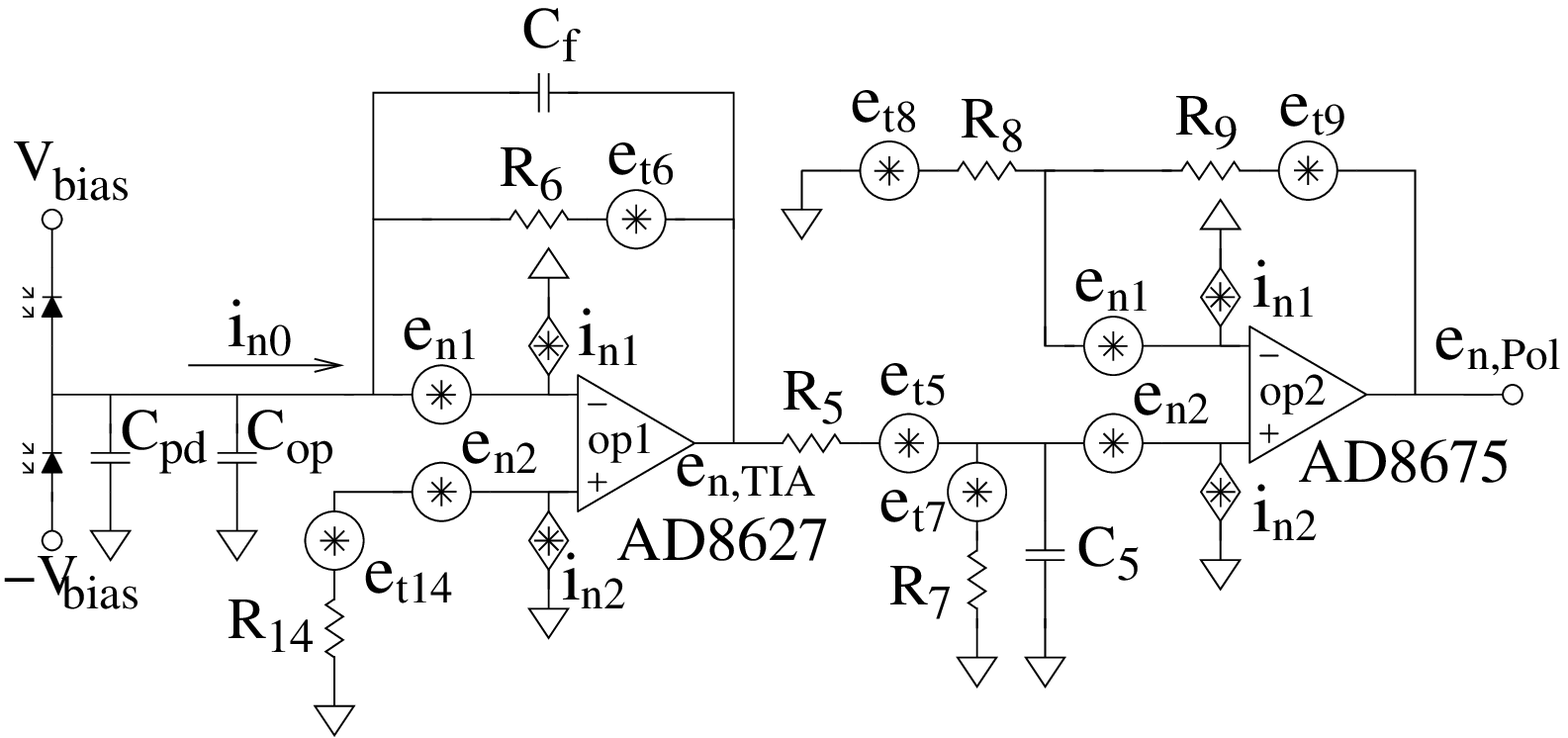}
 \caption {TIA and second amplifier stage implementation including the noise sources considered for the analysis.}
 \label{fig:DAVLLTIA}
\end{figure}

The polarimeter circuit including the same intrinsic noise sources as in Sec.\,\ref{sec:noiseFL} is shown in Fig.\,\ref{fig:DAVLLTIA}. The expected noise for the two stages of the circuit can be computed from Eq.\,\eqref{eq:noisePolarimeter1} and Eq.\,\eqref{eq:noisePolarimeter2}. 

\begin{align}
\label{eq:noisePolarimeter1}
e^ 2_{\rm n,TIA} &\simeq e^{2}_{R_6,t} + i^{2}_{\rm n, op1}|Z_{\rm f}|^2 + \\\nonumber
&+ \left(e^{2}_{\rm n, op1} + e^{2}_{R_{14},t}\right)\frac{1 + (2\pi f R_6C_{\rm T})^2}{1 +  (2\pi f R_6C_{\rm f})^2}\\ 
e^2_{\rm n,Ninv} &= e^{2}_{R_8,t}\frac{R_9^2}{R_8^2} + e^{2}_{R_9,t} + i^{2}_{\rm n, op2}R_9^2 + \left(1+\frac{R_9}{R_8}\right)^2\times  \nonumber\\ 
 &\times \left[e^{2}_{R_7,t} + e^{2}_{R_5,t} + e^{2}_{\rm n, op2} + i^{2}_{n, op2}\left(\frac{R_5R_7}{R_5+R_7}\right)^2\right] \label{eq:noisePolarimeter2}
\end{align}

\noindent where $e_{\rm n, op1}$, $i_{\rm n, op1}$, $e_{\rm n, op2}$, $i_{\rm n, op2}$ are the noise properties of the operational amplifier for the first and second stage, $e_{R,t}$ is the thermal noise in the resistor, $e_{\rm n, TIA}$ is the voltage noise at the output of the TIA, $e_{\rm n, Ninv}$ is the voltage noise at the output of the non-inverting amplifier, $Z_{\rm f}$ is the feedback impedance and $C_{\rm T}$ is the total circuit capacitance, considering feedback capacitance $C_{\rm f} = 7\,{\rm pF}$, op-amp input capacitance $C_{\rm op} = 7.9\,{\rm pF}$ and  photodiode capacitance $C_{\rm pd} = 120\,{\rm pF}$. Noise gain due to the photodiode's shunt resistor ($50\,{\rm M\Omega}$) has been considered negligible, in the same way as the noise contribution due to thermal drifts. In order to ensure loop stability and limit gain peaking or oscillations, the feedback capacitor was chosen large enough to get overcompensation; its drawback is the bandwidth reduction ($\sim 78\,{\rm kHz}$), which is not an issue for the measurement ($\Omega_m= 19.45\,{\rm kHz}$). The sensitivity of the polarimeter is set by $R_6 = 1\,{\rm M\Omega}$, $R_8 = 1.1\,{\rm k\Omega}$ and $R_9 = 11\,{\rm k\Omega}$ and the total output noise (excluding temperature fluctuations which are considered further on in the text) is

\begin{equation}
\label{eq:totalNoisePolarimeter}
e^2_{\rm n,Pol} = e^2_{\rm n,TIA}\left(1+\frac{R_9}{R_8}\right)^2 + e^{2}_{\rm n, Ninv}. 
\end{equation}

\begin{figure}[ht!]
\centering
  \includegraphics[width=1.00\columnwidth]{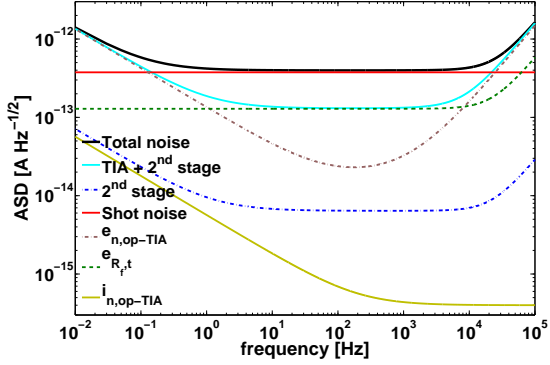}
 \caption {Input noise breakdown for the polarimeter. Shot noise of the light and op-amp voltage noise are the most important noise sources. The black trace is the sum of all the noise sources together.}
 \label{fig:TIAinput}
\end{figure}

\begin{figure}[ht!]
\centering
  \includegraphics[width=1.0\columnwidth]{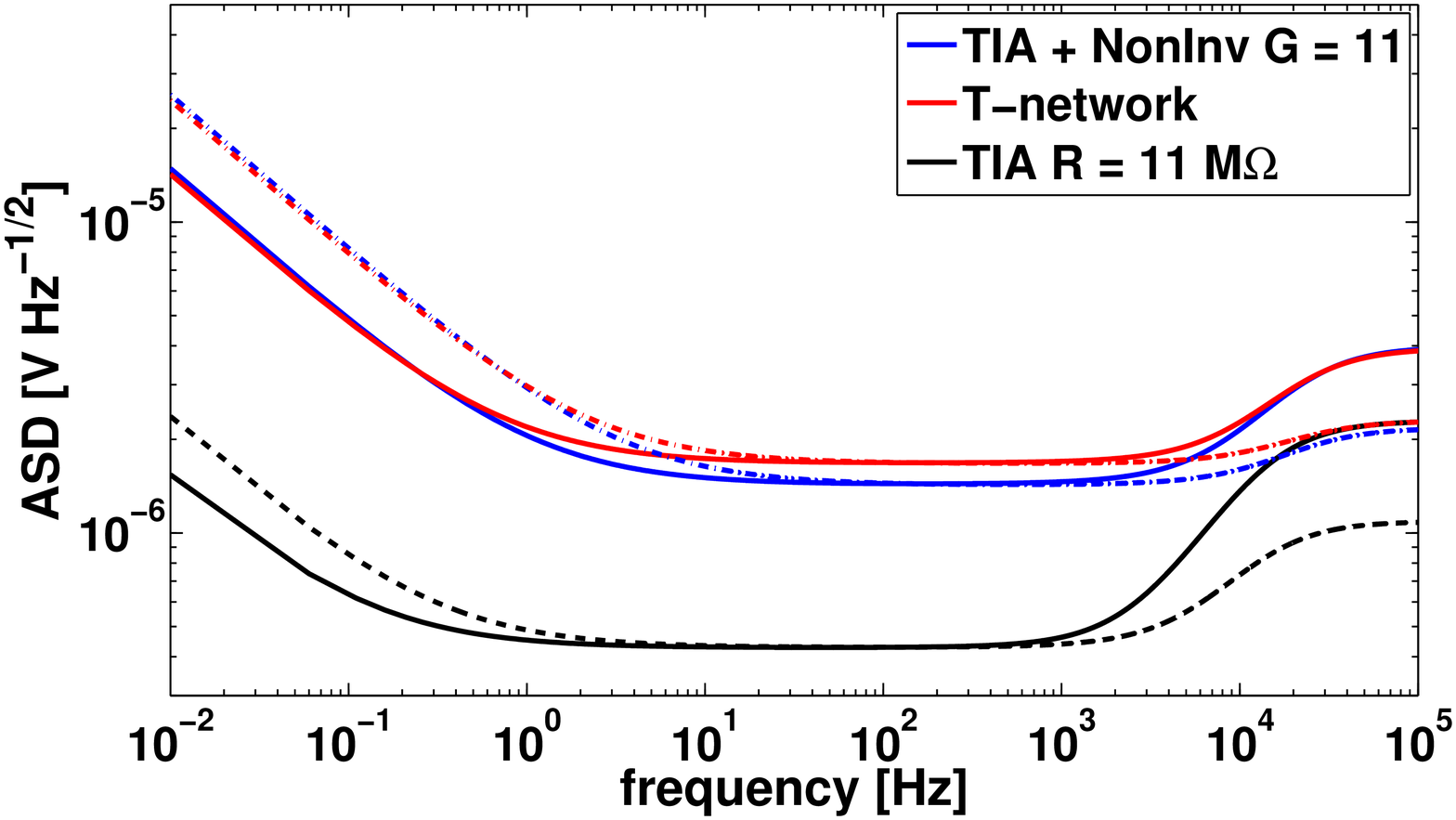}
 \caption {Theoretical amplitude spectral densities for a two-stage current-to-voltage amplifier (Classical TIA +  Non-inverting amplifier), TIA with a T-Network in the feedback loop and TIA with a $11\,{\rm M\Omega}$ feedback resistor. Noise using the AD8627 is displayed as a solid line and the OPA124 as a dashed line.}
 \label{fig:theorpolarimeternoise}
\end{figure}

The polarimeter noise in Fig.\,\ref{fig:TIAinput} was quantified in terms of current spectral density in order to directly compare to the photocurrent shot noise for a $1\,{\rm \mu W}$ beam at 852 nm, which becomes $\sim 0.4\,{\rm pA\,Hz}^{-1/2}$ assuming a silicon photodiode responsivity of $0.55\,{\rm A\,W}^{-1}$. The calculated noise density contributions referred to the input shows that the op-amp voltage noise of the TIA is the main contributor at low frequencies and over $10\,{\rm kHz}$ (see Table\,\ref{tab:componentsPol} for the op-amp characteristics). The high-frequency effect is due to the response of $C_{\rm op} + C_{\rm pd}$, where the gain peaking takes effect and is leveled off by the feedback capacitance $C_{\rm f}$ (Fig.\,\ref{fig:theorpolarimeternoise}, solid red line).  At frequencies between $1\,{\rm Hz}$ and $10\,{\rm kHz}$, the shot-noise level plays the main role in the total spectral noise density, followed by the contribution of the Johnson noise of the large TIA's feedback resistor.

\begin{center}
\begin{table}[ht!]
\caption{Output noise characteristics for the op-amps used in the polarimeter. Corner frequency $f_c$ for the op-amp current noise is  estimated according with the white noise specified by the manufacturer.}
\begin{center}
%\begin{ruledtabular}
\begin{tabular}{c . r . r}
\hline\hline
\multirow{2}{*}{IC} & \multicolumn{1} {c}{$e_n$} & \multicolumn{1} {c}{$f_{c,e_n}$}   & \multicolumn{1} {c}{$i_n$} & \multicolumn{1} {c}{$f_{c,i_n}$}\\
            & \multicolumn{1} {c}{$[{\rm  nV/\sqrt{\rm Hz}}]$} & \multicolumn{1} {c}{$[{\rm Hz}]$} & \multicolumn{1} {c}{$[{\rm pA/\sqrt{\rm Hz}}$]} & \multicolumn{1} {c}{$[{\rm Hz}]$}\\\hline
AD8627  & 17.5 & 59 & 0.004 & 200\\ 
AD8675  & 2.8  & 6 & 0.3 & 2\\  
OPA124  & 8  & 848 & 0.0008 & 1.75\\\hline\hline
\end{tabular}
%\end{ruledtabular}
\end{center}

\label{tab:componentsPol}
\end{table}
\end{center}

As in Sec.\,\ref{sec:noiseFL}, the calculated contribution to the overall temperature dependence of the circuit due to the TC of the resistors is much greater than that due to the drift of the bias/offset current and offset voltage ($\simeq0.08\,{\rm \mu V K^{-1}}$). Therefore, ignoring thermal drifts in the input-referred errors of the operational amplifier, the temperature dependence of the polarimeter  is 

\begin{equation}\label{eq:TCPol}
\alpha_{\rm Pol}(T) = \rho P_{\rm DC}R_6 \sqrt{\alpha_{R6}^2 + 4\alpha_{R,{\rm NI}}^2\left(\frac{R_9}{R_8}\right)^2},
%\alpha_{\rm Pol}(T) = \rho P_{\rm DC}R_6 \sqrt{\alpha_{R6}^2 + \left(2\alpha_{R,{\rm NI}}\frac{R_9}{R_8}\right)^2},
\end{equation}

\noindent where $\rho$ is the photodiode responsivity, $P_{\rm DC}$ is the incident light power and $\alpha_R$ is $25\,{\rm ppm\,K^{-1}}$ and $15\,{\rm ppm\,K^{-1}}$ for the TIA and the non-inverting amplifier,  respectively. Therefore, we obtain that the  polarimeter's TC is $166\,  {\rm \mu V\,K^{-1}}$ ($15\,  {\rm pA\,K^{-1}}$ or $27\,  {\rm pW\,K^{-1}}$). Hence, for the thermal environment in eLISA and even in conventional laboratories ($S_{T,{\rm lab}}^{1/2}< 1\,{\rm mK \,Hz}^{-1/2}$ at $1\,{\rm Hz}$), the  noise contribution due to the TC of the TIA is considered negligible ($<15\,{\rm fA \,Hz}^{-1/2}$).

\begin{figure}[ht!]
\centering
 \includegraphics[width=0.85\columnwidth]{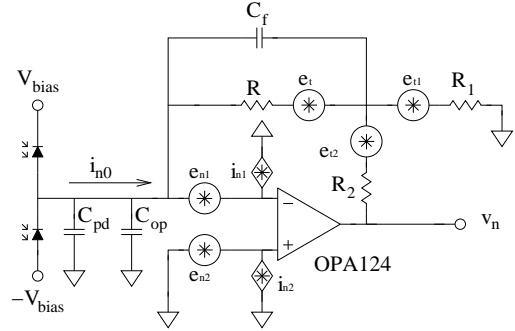}
 \caption {T-network TIA implementation with the addition of the noise sources that originate from the op-amp and resistors.}
 \label{fig:TnetworkTIA}
\end{figure}

In order to reduce the overall noise maintaining the current-to-voltage sensitivity, the second stage could be omitted by increasing the feedback resistor of the TIA. However, for larger values of the feedback resistor, the stray capacitance across the feedback $C_{\rm s}$ has more effect on the bandwidth ($C_{\rm s}\leq 1\,{\rm pF}$ for carefully printed-circuit layout). For that reason, we do not adopt the single TIA option with higher feedback resistor ($11\,{\rm M\Omega}$), though it would be a better option for lower bandwidth applications. As an alternative, a T-network could overcome such drawback, keeping the same value for the largest resistor ($1\,{\rm M\Omega}$) and eliminating the need for the non-inverting amplifier. Fig.\,\ref{fig:TnetworkTIA} shows the circuit for the T-network TIA with the noise sources considered for the analysis. The theoretical voltage noise for this configuration is 

\begin{align}   
 e^2_{\rm n, Tnet} &\simeq  i^{2}_{\rm n, op}|Z_{\rm f}|^2\left|1 + \frac{R_2}{Z_{\rm f}} + \frac{R_2}{R_1}\right|^2 + \left(1 + \frac{R_2}{R_1}\right)^2\times\nonumber\\
 &\times\left(e^{2}_{\rm n, op}\frac{1 + (2\pi f RC_{\rm T})^2}{1 +  (2\pi f RC_{\rm f})^2} + e^{2}_{R{\rm ,t}} + e^{2}_{R_2{\rm ,t}} + e^{2}_{R_1{\rm ,t}}\right). 
 \label{eq:noiseTIAT}
\end{align}

The approach to minimizing the output noise is limited by the restrictions that the resistors selected for the T-network must: 1) hold the same sensitivity and similar measurement bandwidth as the two-stage configuration, and 2) have a maximum resistance value of $1\,{\rm M\Omega}$. The values selected for the T-network are $R = 1\,{\rm M\Omega}$, $R_1 = 48.1\,{\rm k\Omega}$ and $R_2 = 458.8\,{\rm k\Omega}$. The noise spectral densities for the topologies that were analyzed are shown in Fig.\,\ref{fig:theorpolarimeternoise}, where the results exhibit similar noise for the single-stage amplifier with a T-network than for the two-stage TIA. There is also a compromise between high- and low-frequency noise, and some improvements can be achieved from a trade-off between the input voltage noise and the corner frequency of the op-amp. Fig.\,\,\ref{fig:theorpolarimeternoise} shows the performance comparison of the OPA124 and AD8627, where the high-frequency performance is improved at the expense of having higher $1/f$ noise (see Table\,\ref{tab:componentsPol}). The noise has been quantified in terms of output voltage spectral density in order to observe the frequency response of the amplifier's voltage noise contribution. The figure shows the pole response caused by the feedback impedance at the beginning of the high-frequency asymptote, $1 + (C_{\rm pd} + C_{\rm op})/C_{\rm f} $, which is the dominant source of noise over $20\,{\rm kHz}$. For this particular case, the motivation for choosing better photodiodes like the S1223-01 with smaller $C_{\rm pd}$, together with op-amps with lower input voltage noise $e_{\rm n,op}$ would improve the technical noise over $20\,{\rm kHz}$, i.e., at fields $\gtrsim 6\,{\rm \mu T}$.

\begin{figure}[ht!]
\centering
 \includegraphics[width=1.00\columnwidth]{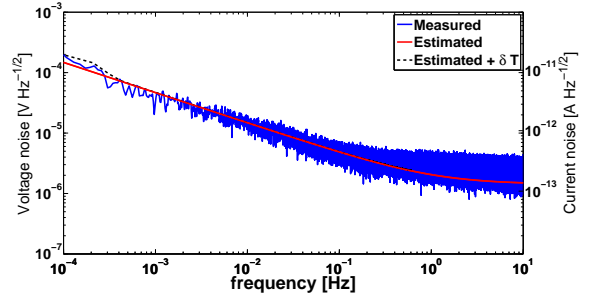}
 \caption {Theoretical and measured noise densities of a two-stage current-to-voltage converter.}
 \label{fig:theorvsMeasPolarimeter}
\end{figure}

For our measurements, low-frequency contribution caused by the polarimeter circuit is less critical since the signal is modulated at frequencies where the $1/f$ behavior is mitigated. As seen in Fig.\,\ref{fig:theorpolarimeternoise}, the corner frequency is around $1.5\,{\rm Hz}$, which would be within the zero-field resonance for small magnetic field measurements ($\sim1\,{\rm nT}$). As a result of the estimated spectral density, the low-frequency contribution is not critical for the possible magnetic field environment in eLISA. In order to confirm the analysis of the circuit, noise measurements were carried out between $0.1\,{\rm mHz}$ and $10\,{\rm Hz}$, where both $1/f$ noise and noise floor are represented. Fig.\,\ref{fig:theorvsMeasPolarimeter} shows that the results are in good agreement with the theoretical predictions using the noise sources considered in the circuit. The measured noise floor is around $2.5\,{\rm  \mu V\,Hz^{-1/2}}$, which corresponds to an equivalent magnetic field noise of $1.5\,{\rm  pT\,Hz^{-1/2}}$. The polarimeter noise was translated to magnetic field according to the characteristics of the magnetic-resonance curve, i.e., the slope of the absorptive curve gives the relation between amplitude and magnetic-resonance frequency. Given that the magnetic-resonance linewidth is $41.6\,{\rm Hz}$ and the peak optical rotation signal is $9.9\,{\rm mV}$, the voltage-to-hertz ratio is $0.476\,{\rm mV/Hz}$.  Another recent magnetometer essentially of the same type but with larger vapor cell and magnetic-resonance linewidth of $2.9\,{\rm Hz}$ reaches a noise level of $50\,{\rm  fT\,Hz^{-1/2}}$ \cite{patton}. With the present design, measurements with a signal-to-noise ratio of $70\,{\rm dB}$ can be made using the lock-in amplifier with an equivalent noise bandwidth (ENBW) of $1.25\,{\rm Hz}$. 

\section{Equivalent magnetic field noise measurements}\label{sec:results}

For low-frequency noise characterization of the magnetometer in the laboratory, different runs during periods of at least twelve hours were carried out. The amplitude spectral density for the magnetic noise measurements of the magnetometer is

\begin{equation}
S_{B,{\rm system}}^{1/2} =  \gamma^{-1}\left(S_{\Omega_m} + S_{\theta,\Omega_m} + 2 C S_{\Omega_L,\Omega_m} S_{\Omega_L,\theta}\right)^{1/2},
\label{eq:noise}
\end{equation}

\noindent where $S_{\Omega_m}$ is the noise power density of the modulation frequency, $S_{\theta,\Omega_m}$ is the noise power density of the phase fluctuations translated to frequency, $\gamma$ is the gyromagnetic ratio for $^{133}$Cs ($3.5\,{\rm Hz\,nT^{-1}}$) and $C$ is the correlation coefficient for partially correlated signals \cite{motchenbacher}.

% or drifts in the duty cycle of the pump laser waveform

\begin{figure}[ht!]
\centering
  \includegraphics[width=1.1\columnwidth]{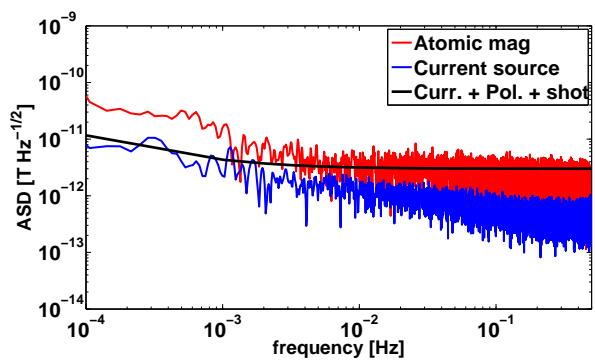}
\caption{Equivalent magnetic field spectral density for the magnetometer (red trace) and equivalent noise applied by the \textit{floating-load} current source (blue trace). Differences at frequencies higher than $10\,{\rm mHz}$ between the experimental measurements and the theoretical behavior of the current source (Fig.\ref{fig:result_floating}) are due to the limited resolution of the digital multimeter (Agilent 34461A). The black trace is the sum of the theoretical noise of the current source, the polarimeter and the shot noise at $\Omega_m= 19.45\,{\rm kHz}$.}
\label{fig:asdAtomicMag}
\end{figure}

The measured equivalent magnetic field noise of the system shown in Fig.\,\ref{fig:asdAtomicMag} is around $50\,{\rm pT\,Hz}^{-1/2}$ at $0.1\,{\rm mHz}$ and fulfills the requirement given in Eq.\,\eqref{eq:req}. The current applied to the coil has also been measured and converted to the equivalent magnetic noise, where as expected from the estimated value in Sec.\,\ref{sec:noiseFL}, the noise level is around $8\,{\rm pT\,Hz}^{-1/2}$ at 0.1 mHz. The excess noise observed in the magnetometer at sub-millihertz frequencies is well over the noise applied by the coil. Then, we prove that the characterized electronic noise contribution does not limit the performance of the magnetometer at the lower end of the eLISA bandwidth. Now that the main electronic noise sources have been characterized (current source and detector noise), further work can be done in order to unveil the noise limits at low-frequency, such as light shifts induced by the laser light or alkali density fluctuations. At higher frequencies, the magnetometer noise-floor measurement is $\sim 2.5\,{\rm pT\,Hz^{-1/2}}$, which is in agreement with the equivalent magnetic field noise calculated for the polarimeter in Sec.\,\ref{sec:NoisePol} (the theoretical estimation considering the polarimeter and the shot noise for $\Omega_m= 19.45\,{\rm kHz}$ is $3\,{\rm pT\,Hz^{-1/2}}$). The main contributors to the noise floor are the op-amp voltage noise for the first stage of the TIA and the photocurrent shot-noise of the incident light. The theoretical estimation of the additive noise due to the polarimeter, shot-noise, and current source is shown in Fig.\,\ref{fig:asdAtomicMag} (black trace). 

%it appears that the equivalent noise will be dominated by the intrinsic noise of the magnetometer or by the residual magnetic field drifts %inside the shielding, which could be caused by the thermal and magnetic variations in the laboratory.

%, which could be caused by the thermal and magnetic variations in the laboratory

%It appears that the equivalent noise will be dominated by the intrinsic noise of the magnetometer or by the residual magnetic field drifts inside the shielding.

\section{Conclusion}\label{sec:conclusions}

We presented the noise performance of an atomic magnetometer based on NMOR with modulated light at sub-millihertz frequencies. To quantify and discern electronic contributions from the overall noise measurement, the polarimeter circuit and the current source for the leading field were characterized in terms of their intrinsic noise and thermal dependence. The estimation of the noise of the circuits is in good agreement with the measurements, which are clearly dominated by the electronic $1/f$ noise at lower frequencies and where, the thermal effects start to appear below the measurement bandwidth ($< 0.1\,{\rm mHz}$). The designed current source creates a quiet magnetic environment which allows us to measure the atom-based sources of drift within the desired bandwidth. In addition, the polarimeter circuit operates below the photon shot-noise level between $1$ and $10\,{\rm kHz}$; above this frequency range the current-to-voltage converter exhibits gain peaking, which can be readily improved.   The magnetometry technique presented in this paper proves to be a promising technology for eLISA in terms of sensitivity, given the fact  that it is  well within the eLISA requirement at 0.1 mHz. However, due to the size, weight and power restrictions for space applications, further work on sensor miniaturization and its effects on the sensitivity and the low-frequency behaviour should be performed. This work might be also useful in other applications beyond the scope of eLISA, where small sensors with long-term stability are required.

\section*{Acknowledgments}
Support for this work came from Project AYA2010-15709 of the Spanish Ministry of Science and Innovation (MICINN), ESP2013-47637-P of the Spanish Ministry of Economy and Competitiveness (MINECO), and 2009-SGR-935 (AGAUR). I. M. is very grateful to the members of the Budker Group for their help and hospitality during his stay at the University of California at Berkeley.

\end{document}